# A new charge reconstruction algorithm for the DAMPE silicon microstrip detector

Rui Qiao[1], Wen-Xi Peng[1;1)], Dong-Ya Guo[1], Hao Zhao[1], Huan-Yu Wang[1], Ke Gong[1], Fei Zhang[1]

[1] Institute of High Energy Physics, Chinese Academy of Sciences, Beijing 100049, China

**Abstract**: The DArk Matter Particle Explorer (DAMPE) is one of the four satellites within the Strategic Pioneer Research Program in Space Science of the Chinese Academy of Science (CAS). The Silicon-Tungsten Tracker (STK), which is composed of 768 singled-sided silicon microstrip detectors, is one of the four subdetectors in DAMPE, providing track reconstruction and charge identification for relativistic charged particles. The charge response of DAMPE silicon microstrip detectors is complicated, depending on the incident angle and impact position. A new charge reconstruction algorithm for the DAMPE silicon microstrip detector is introduced in this paper. This algorithm can correct the complicated charge response, and was proved applicable by the ion test beam.

**Keywords**: DAMPE; STK; Silicon microstrip detector; Charge reconstruction; Charges sharing.

**PACS**: 29.40.Mc, 95.55.-n

## 1 Introduction

The DArk Matter Particle Explorer (DAMPE) has been successfully launched on December 17, 2015. DAMPE can detect electrons and photons in 5GeV ~ 10 TeV in order to identify possible dark matter signatures. DAMPE can also measure the charged cosmic ray fluxes in 100 GeV ~ 500 TeV which brings understanding to the production, acceleration and propagation of the cosmic rays [1].

DAMPE is made up of 4 subdetectors: plastic scintillator detectors (PSD), a silicon-tungsten tracker-converter (STK), a Bismuth Germanium Oxide imaging calorimeter (BGO) and neutron detectors (NUD), as shown in Fig. 1. The PSD consists of 2 orthogonal layers of scintillator bars, covering an area of $82 \times 82$ cm². PSD can provide electron/gamma separation and charge identification up to Argon with resolution around 0.25 charge unit. The STK consists of 12 orthogonal layers of single-sided silicon microstrip detectors and 3 layers of 1mm thick tungsten plates. The angular resolution is around 0.10 degrees for 100GeV photons and the charge resolution is better than 0.10 charge unit for protons. The BGO calorimeter is made of 14 layers of BGO crystal bars with a total thickness equivalent to 31 radiation lengths. Electrons and photons from 5GeV to 10TeV can be covered by the BGO. BGO also provide triggers for the other 3 payloads. The NUD consists of 4 boron-doped plastic scintillators with a dimension of $30 \times 30 \times 1$ cm³ each. Neutrons from the hadron-induced showers can be detected by the NUD. This can improve the electron/proton separation power above 100 GeV.

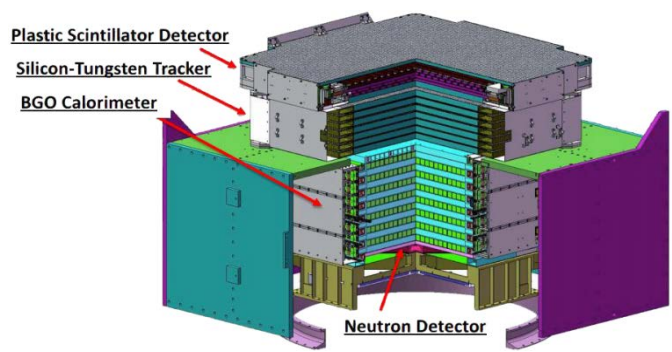

Fig. 1. Schematics of the DAMPE detectors.

## 2 The DAMPE silicon microstrip detectors

The core of the STK is 768 AC-coupled single-sided silicon microstrip detectors with a dimension of $95 \times 95 \times$

* Supported by National Natural Science Foundation of China (U1738133), State's Key Project of Research and Development Plan (2016YFA0400204), Strategic Pioneer Research Program in Space Science of the Chinese Academy of Science, Youth Innovation Promotion Association, CAS.
1) E-mail: pengwx@ihep.ac.cn, qiaorui@ihep.ac.cn

$0.32 \text{ mm}^3$ each. The junction side of the detector is segmented into 768 p+ implantation strips with a pitch of 121 μm and width of 48 μm. Every four silicon microstrip detectors are daisy chained together to form a basic unit named ladder. These four silicon microstrip detectors share the same readout electronics and bias voltage. This helps to reduce the number of electronics and the power consumptions and is widely used in space projects [2,3,4,5] and ground experiments [6,7]. For each silicon microstrip detector, only half of the implantation strips (384 strips) are AC coupled to the front-end electronics and amplified by six 64-channels VA140 ASICs [8]. These 384 implantation strips are named readout-strips. The other 384 implantation strips are not AC coupled to the front-end electronics and are named float-strips. This float-strips design inherit from the other silicon microstrip detectors [2,4,5] and can improve the spatial resolutions with limit electronics.

However, this float-strips design has a drawback of charge collection efficiency diversities. The charge collection efficiency of the float-strip incidence is less than that of the readout-strip incidence. This difference of charge collection efficiency depends on the sensor layout [9], and reaches around 25% for AGILE [10] and AMS-02 [11].

According to the Bethe-Bloch formula [12], the deposited energy of a relativistic charged particle is proportional to $Z^2$ and the length of trajectory inside the detector. Electron-hole pairs are excited by the deposited energy inside the silicon detector, and are collected by the implantation strips. Then they are processed by the electronics and finally digitalized as the amplitudes of consecutive readout channels. These consecutive readout channels are usually referred to as a cluster. The cluster amplitude, defined as the total amplitudes of all the channels within this cluster, is usually characterized as an estimator of the deposited energy. After trajectory corrections, the deposited energy is proportional to $Z^2$ for relativistic charged particles. As a result, the cluster amplitude after trajectory corrections is usually characterized as an estimator of $Z^2$.

Fig. 2 shows the cluster amplitudes of normal incident carbons as a function of $\eta$ for the DAMPE silicon microstrip detectors. The variable $\eta$ is an estimator of the impact position. $\eta$ is defined as the Center-of-Gravity, in units of the readout-strips pitch, of the two consecutive readout-strips with the maximum amplitudes [10,11]. $\eta$ close to 0 and 1 correspond to the readout-strip incidence, while $\eta$ close to 0.5 correspond to the float-strip incidence. The black circles represent the most probable value (MPV) of the Landau distribution convoluted with a Gaussian-distributed noise, and the error bars correspond to the standard deviations. The difference of MPVs between the float-strip incidence ($\eta{\sim}0.5$) and the readout-strip incidence ($\eta{\sim}0$ or 1) is around 35%. The above situation is changed for large incident angles as more than one implantation strips are fired. The charge collection efficiency is contributed by both the fired readout-strips and the float-strips. As a result, the charge collection efficiencies of normal incidence are different from those with large incident angles.

To summarize, the charge response of the DAMPE silicon microstrip detectors is complicated. Their cluster amplitude not only depends on the $Z$, but also depends on the impact position and the inclination angle. Similar phenomenon has also been reported by AGILE [10] and AMS-02 [13,14].

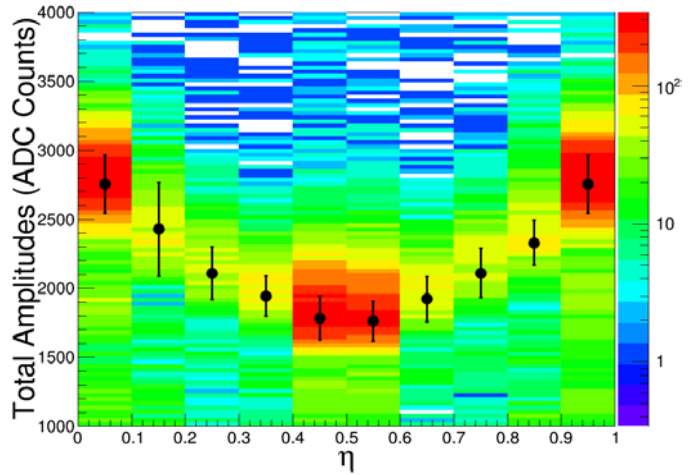

Fig. 2. Position dependency for normal incident carbons

# 3    Charge reconstruction algorithms

## 3.1  The Binning Correction Algorithm

It is necessary to apply some corrections to the cluster amplitude. A commonly used correction algorithm is to multiply different correction factors for different impact positions and different angles [13,14].

First, the impact positions and the incident angles are divided into several bins. For each bin, a cluster amplitude spectrum is accumulated from the clusters with known $Z$. Then these charge spectra are fitted and the mean values are evaluated. To correct the complicated response, the mean values of all bins should be equalized to the same value. The equalization constant of each bin is the correction factor.

To apply the correction for a cluster, the corresponding bin (the impact position and incident angle) of this cluster is evaluated. The cluster amplitude will multiply the correction factor of this bin. After the correction, the cluster amplitude is a better estimator of $Z^2$.

This correction algorithm will be referred to as the "Binning Correction Algorithm" in this paper. This algorithm has several disadvantages:

- Sufficient statistics are required to accurately calibrate the correction factors, as the statistics are divided into several bins (e.g. 361 bins in ref. 14). However, the statistics for large incident angles are usually limited by the trigger efficiency. So it is a great challenge to achieve accurate correction factors for large incident angles.

- Each cluster amplitude is corrected according to its impact position and inclination angle. If one of these two parameters (especially the impact position) is incorrect, the reconstructed charge will be biased without being noticed.

### 3.2 The charges sharing algorithm

To overcome the limits of the Binning Correction Algorithm described above, a new charge reconstruction algorithm for the DAMPE tracker is introduced. This algorithm is based on the linear charges sharing assumption, and will be referred to as the "Charges Sharing Algorithm" in the following discussions.

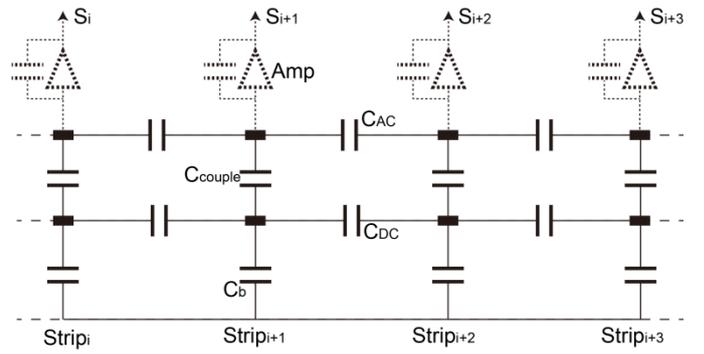

Fig. 3. Equivalent model of the silicon microstrip detector

Silicon microstrip detector are usually model as a network of capacitors [15,16] as shown in Fig. 3. The $C_b$, $C_{couple}$, $C_{DC}$, $C_{AC}$ are the strip-to-backplane capacitor, the coupling capacitor, the interstrip capacitor between the implantation strips and the aluminum electrodes, respectively. The charges are finally collected by the charge-sensitive pre-amplifiers on top of Fig. 3. Only the readout-strips of DAMPE silicon microstrip detectors are connected to the pre-amplifiers, while the float-strips have no connections. However, the strips in Fig. 3 are not specified as the readout-strips or the float-strips. So the pre-amplifiers in Fig. 3 are plotted in dashed lines.

If the pre-amplifiers have ideal response, the pre-amplifier output signal can be treat as the total contributions from individual strips:

$$S_i\left(Q_m, Q_{m+1}, ..., Q_{n-1}, Q_n\right) = \sum_{k=m}^{n} S_i\left(Q_k\right) \qquad (1)$$

$S_i$ is the pre-amplifier amplitude of the $i$-th strip. $Q_k$ is the collected charges of the $k$-th strip. The left side of (1) is the pre-amplifier amplitude with multiple strips are fired. This usually happen with inclined incident particles. Each element on the right side of (1) is the pre-amplifier amplitude contributed by only one fired strip. This is usually caused by normal incident particles.

The charges collection time in the DAMPE silicon sensors is around 35 ns, while the integration time of the pre-amplifier is 6.5 μs. As the integration time is much longer than the charges collection time, the pre-amplifier amplitude with single fired strip is assumed to be proportional to the collected charges:

$$S_i(Q_k) = G_i R_{i,k} Q_k \qquad (2)$$

$G_i$ is the gain of the $i$-th pre-amplifier, and will be equalized before charge reconstruction. $R_{i,k}$ is the charges sharing ratio from the $k$-th strip to the $i$-th pre-amplifier. As silicon microstrip detectors are usually manufactured with good equalization, the charges sharing ratio $R_{i,k}$ can be simplified as:

$$R_{i,k} = R_{|i-k|} \qquad (3)$$

$R_{|i-k|}$ is the simplified charges sharing ratio, which only depends on the distance between the $i$-th pre-amplifier and the $k$-th strip.

Considering an inclined incident charged particle excites charges from the $m$-th strip to the $n$-th strip, these charges can be expressed in a matrix $Q_{(n-m+1)\times1}$. Each element of this matrix is the charges collected by the corresponding strip. The summation of all the elements in this matrix, defined as $Q_{total}$, is the total excited charges which is proportional to the deposited energy. To be mentioned, this $Q_{total}$ is not affected by the complicated charge response of the DAMPE silicon microstrip detectors. After trajectory corrections, the $Q_{total}$ is a good estimator of $Z^2$.

These excited charges will induce a cluster, which is composed of the signals from the $k$-th pre-amplifier to the $l$-th pre-amplifier. With the gain of pre-amplifiers equalized, the information of this cluster can be expressed in a matrix $S_{(l-k+1)\times1}$. Each element of this matrix is the equalized amplitude of the corresponding channels in the cluster.

According to the formulas (1)~(3), we obtain:

$$\frac{1}{G}S_{(l-k+1)\times1} = R_{(l-k+1)\times(n-m+1)} \cdot Q_{(n-m+1)\times1}$$
$$r_{i,j} = R_{|l+i-m-j|} \qquad (4)$$

$G$ is the equalized gain of pre-amplifiers, $R_{(l-k+1)\times(n-m+1)}$ is the matrix of the charges sharing ratios, and $r_{i,j}$ is the elements of this matrix. $R_{|l+i-m-j|}$ is the charges sharing ratio between the $(l+i)$-th pre-amplifier and the $(m+j)$-th strip, or the charges sharing ratio with the same distance (see formula (3)).

In formula (4), the matrix $S_{(l-k+1)\times1}$ is already known. The gain of pre-amplifiers $G$ and the matrix $R_{(l-k+1)\times(n-m+1)}$ can be calibrated as discussed in section 4.2. The matrix $Q_{(n-m+1)\times1}$ or its summation $Q_{total}$ is to be determined. By inversing formula (4), we obtain:

$$Q_{(n-m+1)\times1} = \frac{1}{G}R_{(l-k+1)\times(n-m+1)}^{-1} \cdot S_{(l-k+1)\times1} \qquad (5)$$

$R_{(l-k+1)\times(n-m+1)}^{-1}$ is the inverse matrix of $R_{(l-k+1)\times(n-m+1)}$. If formula (5) is not underdetermined, which usually happens for non-float-strip design silicon microstrip detectors, the matrix $Q_{(n-m+1)\times1}$ or its summation $Q_{total}$ can be solved analytically.

For float-strip design detectors, such as the DAMPE silicon microstrip detectors, formula (5) is usually underdetermined. Some approximations are needed before solving formula (5).

Electron-hole pairs are generated along the trajectory, and then they are drift under the influence of electric field and finally collected by the corresponding electrodes. The charge diffusion during the drift time is around 9 μm, which can be ignored compared to the implantation pitch (121 μm) of the DAMPE silicon microstrip detectors. As a result, we assume a uniform distribution of electron-hole pairs along the trajectory, and these charges are collected only by the corresponding strips, as shown in Fig. 4.

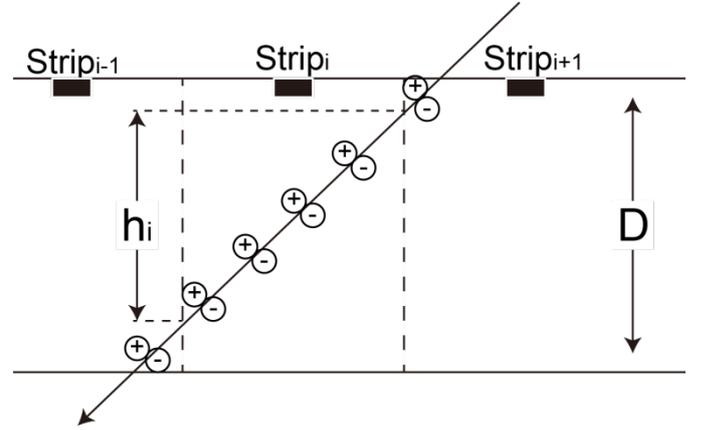

Fig. 4. Simplified model of charges collection in silicon microstrip detectors

The black rectangles stand for the implantation strips. The vertical dashed lines are assumed to be the boundaries between strips, and the charges within the boundaries are assumed to be collected by the corresponding strip. The arrow is the trajectory of incident charged particles, and is fully defined by the impact position and the incident angle. The collected charges of the $i$-th strip can be simplified as:

$$Q_i = Q_{Total} \frac{h_i}{D} \qquad (6)$$

where $h_i$ is the vertical projection of the trajectory within the boundaries of the $i$-th strip, as shown in Fig. 4. $D$ is the thickness of the detector.

In formula (6), $Q_{total}$ is to be determined, and $h_i$ is determined by the impact position and the incident angle. The DAMPE tracker has a good angular resolution (0.10 degrees for 100 GeV photons), and the incident angle can be considered as known. As each element in the matrix $Q_{(n-m+1)\times 1}$ can be easily calculated via formula (6), there are only two variables to be determined in formula (5): $Q_{total}$ and the impact position. As a result, the total excited charges $Q_{total}$ can be calculated analytically via formula (5), as long as the cluster consists of at least two channels. After the trajectory corrections, the solved $Q_{total}$ is an estimator of $Z^2$.

The Charges Sharing Algorithm has several advantages compared to the Binning Correction Algorithm:

- The charges sharing ratios have no relationship with the incident angle. So we can use the events with small incident angles to calibrate the charges sharing ratios. Then the ion charges with large incident angles, which have low statistics limited by the trigger, can be reconstructed using the same charges sharing ratios.
- If formula (5) is overdetermined, residues can be calculated. A large residue implies a poor charge reconstruction quality.

## 4    The ion test beam and data analysis

### 4.1  Experimental set-up

The ion test beam was carried out at CERN SPS (Super Proton synchrotron) in November 2015 in order to investigate the charge measurement behavior of STK ladders. The layout of the test beam is shown in Fig. 5. The primary beam was 60 GeV/n Lead and the secondary beams were selected by the magnet with $A/Z = 2$. Eight flight-model STK ladders were installed perpendicular to the beam in a parallel direction. Secondary fragments were identified by two plastic scintillators which were installed upstream and downstream of the STK ladders respectively. Twelve orthogonal single-sided silicon microstrip detectors (SSDs) were installed between the plastic scintillators and the STK ladders, in order to study the effects of multiple-scattering from tungsten converters and the support structure. The STK ladders were readout, digitalized by the flight-model Tracker Readout Boards [17]. The pedestal calibration, pedestal removal, cluster finding and data compression were also processed by the Tracker Readout Boards [18].

At the very end of the experiment, the eight STK ladders were tilted to nine degrees for a very short period, in order to study the charge reconstruction for inclined incident ions. However, the statistics of nine degrees inclination were so limited. It was hard to reconstruct charges using the Binning Correction Algorithm. So it was a good chance to check the capability of the Charges Sharing Algorithm towards low statistics inclined tracks.

In ref. 19, the particle identification of the plastic scintillators and the gain equalization of pre-amplifiers had been introduced. The charge reconstruction for normal incident particles using the Binning Correction Algorithm had also been reported in ref. 19. This paper will focus on the charge reconstruction using the Charges Sharing Algorithm.

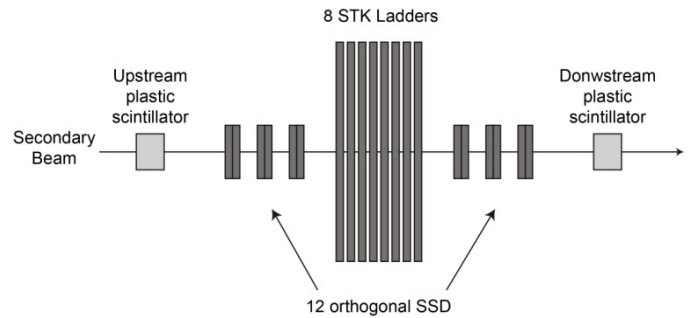

Fig. 5. Layout of the test beam.

### 4.2  Calibration of the charges sharing ratio

The charges sharing ratios defined in (3) were calibrated using the normal incident beams.

First of all, we selected carbons ($Z$=6) from the plastic scintillators for calibration. This was due to the electronics saturation of STK ladders. Ions heavier than Carbon with readout-strip incidence suffer from strong non-linearity. Among the events with linear response, carbons had the highest statistics compared to berylliums and borons.

Then we selected the events with single incident strip. The event selections were accomplished by the $\eta$, as shown in Fig. 6. The readout-strip incidence were defined by $\eta<0.1$ and $\eta>0.9$, while the float-strip incidence were defined by $0.4<\eta<0.6$.

The positions of the float-strips can be defined as 1, 3, 5 …, 767 in the unit of the implantation pitch, while the positions of the readout-strips are defined as 2, 4, 6 …, 768. As explained in section 3.2, the charges sharing ratios are related to the distance between the pre-amplifiers and the implantation strips. For the readout-strip incidence, this distance is always an even number in the unit of implantation pitch, as pre-amplifiers are always connected with the readout-strips. That means the readout-strip incidence can be used to calibrate the charges sharing ratios $R_0$, $R_2$, etc. Similarly, the float-strip incidence can be used to calibrate the charges sharing ratios $R_1$, $R_3$, etc.

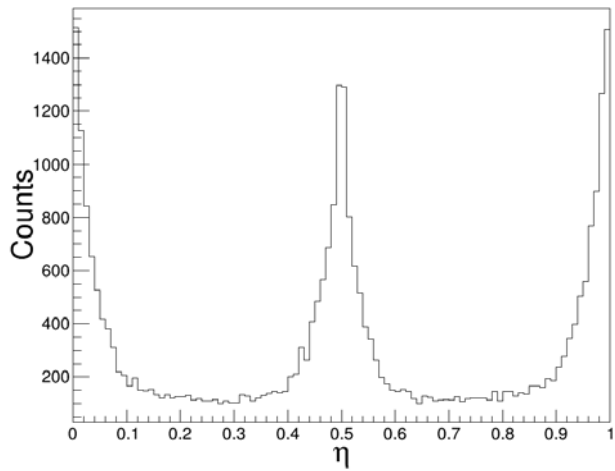

Fig. 6. Spectrum of $\eta$ for normal incident carbons

To calibrate the charges sharing ratios, the amplitude spectra from the corresponding channels were accumulated. Fig. 7 shows an example to calibrate the ratios $R_0$ and $R_1$. Fig. 7a was accumulated during the readout-strip incidence by the amplitude of the same incident readout-strip. As the distance (between the incident readout-strip and itself) is always 0, Fig. 7a was used to calibrate the ratio $R_0$. Fig. 7b was accumulated during the float-strip incidence by the total amplitudes of its two neighboring readout-strips. As the distances between the incident float-strip and its two neighboring readout-strips are always 1, Fig. 7b was used to calibrate the ratio $R_1$.

These two spectra were fitted using the Landau distribution convoluted with a Gaussian distributed noise. The most probable values (MPVs) of these two spectra correspond to $G \times 6^2 \times R_0$ and $G \times 6^2 \times 2R_1$. $G$ is the equalized gain of pre-amplifiers, 6 is the charge of incident carbons and 2 is attributed to the 'two' neighboring readout-strips.

The absolute value of $G$ is hard to calibrate, as the pre-amplifier amplitudes induced by charged particles are always affected by the charges sharing ratios. For simplicity, we take $R_0$ to be 100% and $G$ can be calibrated as 74.417 ADC counts per square charge unit. The charges sharing ratio $R_1$ is 28.7% according to the MPV of Fig. 7b.

The charges sharing ratios with the distance greater than 1 are much smaller than $R_0$ and $R_1$. For example, $R_2$ is around 0.5%.

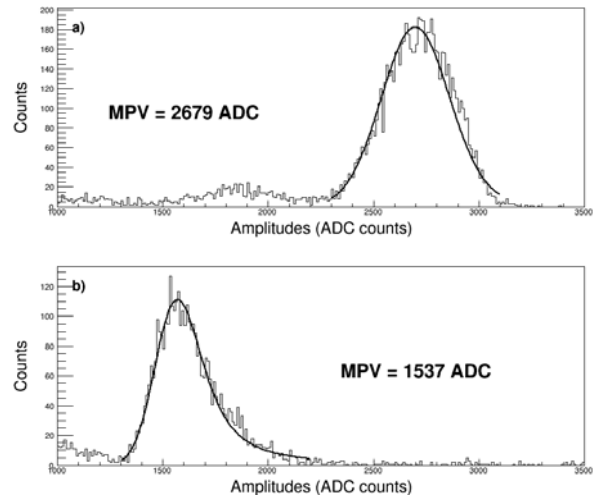

Fig. 7. Spectra to calibrate the charges sharing ratio $R_0$ (top) and $2 \times R_1$ (bottom)

## 4.3 Cluster selections

As the Charges Sharing Algorithm relies on several assumptions, cluster selections should be processed before the charge reconstruction:

- In general, at least two channels were required within a cluster, because there are two variables ($Q_{total}$ and the impact position) to be determined in formula (5). The single channel clusters were mostly attributed to $Z=1$ particles hitting the readout-strips. For these clusters, the impact position was fixed as the only channel, and $Q_{total}$ could be solved via formula (5).

- Clusters should not suffer from strong electronics saturation, as the Charges Sharing Algorithm is a linear algorithm. These clusters were identified by their maximum channel amplitude smaller than 3000 ADC counts [19]. For readout-strip incidence, ions up to Carbon ($Z=6$) could be reconstructed. For float-strip incidence, ions up to Neon ($Z=10$) could be reconstructed because of the loss of charges collection efficiency.

### 4.4 Consistency check

It is necessary to check the consistency between the Binning Correction Algorithm and the Charges Sharing Algorithm, because the Binning Correction Algorithm has been used in previous experiments. The consistency was checked using the normal incident beams, as the statistics of nine degrees inclination were too limited.

A set of correction factors has already been calibrated using the Binning Correction Algorithm [19]. The correction factors and the errors, from Boron ($Z=5$) to Oxygen ($Z=8$), are shown in Fig. 8 as a function of $\eta$.

Given the impact position and $Q_{total}$ set to 1, the information of the induced cluster, expressed in the matrix $S_{(l-k+1)\times 1}$, was well defined by formula (4). Each element in the matrix $Q_{(n-m+1)\times 1}$ can be calculated according to formula (6). The gain of pre-amplifiers and the charge sharing ratios had already been calibrated, as discussed in section 4.2. With the information of this cluster, the corresponding correction factor and $\eta$ can be evaluated. By changing the impact position, the relationship between the correction factors and $\eta$ is deduced, and is plotted in the dashed curve in Fig. 8.

As shown in Fig. 8, the deduced correction factors from the Charges Sharing Algorithm match those from the Binning Correction Algorithm within the range of errors.

This proves the consistency between the Binning Correction Algorithm and the Charges Sharing Algorithm.

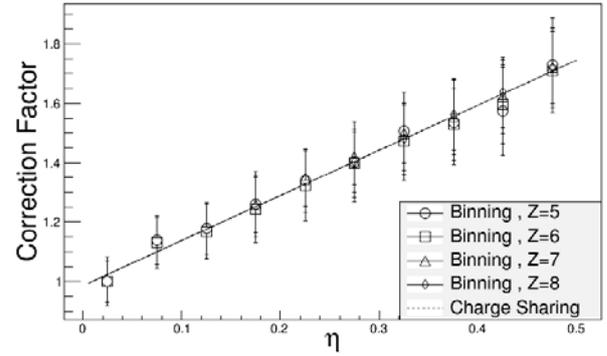

Fig. 8. Comparison of correction factors for normal incident beams

### 4.5 Charge reconstruction using the Charges Sharing Algorithm

With the charges sharing ratios and the gain of pre-amplifiers calibrated in section 4.2, the total excited charges $Q_{total}$ was calculated analytically using the formulas (5) and (6). $Q_{total}$ was in the unit of the average total excited charges from normal incident protons. $\sqrt{\cos(\theta) \cdot Q_{Total}}$ was characterized as the reconstructed charge, where $\cos(\theta)$ was the trajectory correction.

The reconstructed charge spectrum of nine degrees incident carbons is shown in Fig. 9. The charge peak is well distinguished in spite of low statistics. It was fitted using the Landau distribution convoluted with a Gaussian noise, and the sigma was 0.21 charge unit.

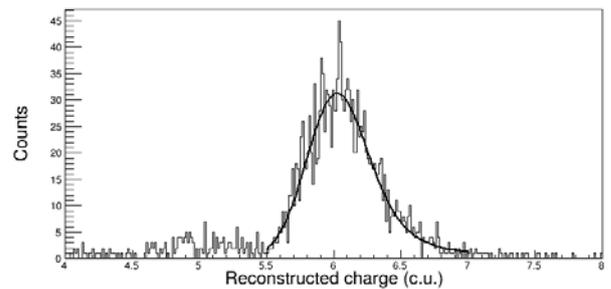

Fig. 9. Charge spectrum of nine degrees incident Carbons using the Charges Sharing Algorithm

The comparison of charge resolutions between the Binning Correction Algorithm and the Charges Sharing

Algorithm is shown in Fig. 10. The resolutions of normal incidence using the Charges Sharing Algorithm were similar to the Binning Correction Algorithm. This is not surprising because of the consistency discussed in section 4.4. The ion charges of nine degrees inclination were also well reconstructed, which proves the applicability of the Charges Sharing Algorithm towards low statistics inclined incidence. The sigma of Beryllium ($Z$=4) of nine degrees inclination is not presented in Fig. 8 due to the extremely low statistics.

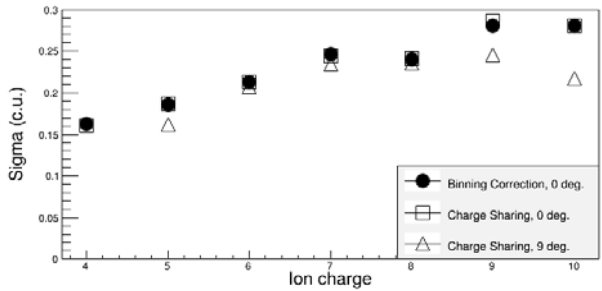

Fig. 10. Charge resolution comparison between the Binning Correction Algorithm and the Charges Sharing Algorithm

## 5    Conclusion

The DAMPE silicon microstrip detectors have complicated charge response due to the float-strip design. The cluster amplitudes are affected by the inclination angle and the impact position, and can be corrected using the commonly used Binning Correction Algorithm. However, it is hard to achieve accurate correction factors for low statistics, especially the events of large incident angles with low trigger efficiencies. What's more, the reconstructed charges may be biased due to the application of inaccurate parameters without being noticed.

A new algorithm, named the Charges Sharing Algorithm, is introduced to reconstruct charges using a linear algorithm. The Charges Sharing Algorithm relies on a set of charges sharing ratios, which can be calibrated using the events with small incident angles. Then the ion charges, including those with large incident angles, can be easily reconstructed with the same set of charges sharing ratios. If the cluster size is greater than two, the residues may be used to evaluate the quality of charge reconstruction.

The charge reconstruction was studied using $A/Z$ =2 fragments from 60GeV/n Lead primary beams at CERN SPS. The normal incident beams were used to calibrate the charges sharing ratios. The consistency between the Binning Correction Algorithm and the Charges Sharing Algorithm was confirmed. The charge resolutions using the Binning Correction Algorithm and the Charges Sharing Algorithm were similar. The nine degrees incident beams were used to check the applicability of the Charges Sharing Algorithm towards low statistics inclined incidence. The ion charges of nine degrees incidence were reconstructed with good resolutions.

## References

1   J. Chang, G. Ambrosi, Q. An, et al., Astroparticle Physics, 95: 6, (2017)

2   P. Picozza, A.M. Galper, G. Castellini, et. al., Astroparticle Physics, 27:296-315, (2007)

3   W. B. Atwood, A. A. Abdo, M. Ackermann, et. al., The Astrophysical Journal, 697:1071–1102, (2009)

4   M. Prest, G. Barbiellini, G. Bordignon, et. al., Nuclear Instruments and Algorithms in Physics Research A 501: 280–287 , (2003)

5   P. Zuccon, Nuclear Instruments and Algorithms in Physics Research A 596: 74–78, (2008)

6   D. Abbaneo, Nuclear Instruments and Algorithms in Physics Research A 518: 331–335, (2004)

7   A. Ahmad, Z. Albrechtskirchinger, P.P. Allport, Nuclear Instruments and Algorithms in Physics Research A 578: 98–118, (2007)

8   Integrated Detector Electronics AS, (http://www.ideas.no)

9   M. Krammer, H. Pernegger, Nuclear Instruments and Algorithms in Physics Research A 397: 232–242, (1997)

10  G. Barbiellini, G. Fedel, F. Liello, et. al., Nuclear Instruments and Algorithms in Physics Research A 490: 146–158 (2002)

11  J. Alcaraz, B. Alpat, G. Ambrosi, et. al., Nuclear Instruments and Algorithms in Physics Research A 593: 376–398 (2008)

12  D.E. Groom et al., The Review of Particle Physics. The European Physical Journal, C **15**(1):1, (2000)

13  B. Alpat, G. Ambrosi, Ph. Azzarello, et. al., Nuclear Instruments and Algorithms in Physics Research A 540: 121–130, (2005).

14  P. Saouter, Nuclei identification with the AMS-02 Silicon Tracker, IEEE Nuclear Science Symposium and Medical Imaging Conference (NSS/MIC) : (2014).

15  M. A. Frautschi, M. R. Hoeferkamp, S. C. Seidel, Nuclear Instruments and Algorithms in Physics Research A 378: 284–296, (1996).

16  N. Bacchetta, D. Bisello, C. Calgarotto, et. al., IEEE Transactions on Nuclear Science 43(3): 1213–1219, (1996).

17  Fei Zhang, Wen-Xi Peng, Ke Gong, et. al., Chinese Physics C 40(11): 116101, (2016)

18  Yi-Fan Dong, Fei Zhang, Rui Qiao, et. al., Chinese Physics C 39(11): 116202, (2015)

19  Rui Qiao, Wen-Xi Peng, Dong-Ya Guo, et. al., arXiv: 1705.09791